\documentclass[journal,twocolumn]{IEEEtran}
\usepackage{amsfonts}
\usepackage{times}
\usepackage{graphicx}
\usepackage{latexsym}
\usepackage{dsfont}
\usepackage{amssymb}
\usepackage{amsmath}
\usepackage{cite}
\usepackage{verbatim}
\usepackage{subfigure}

\newcommand{\figref}[1]{{Fig.}~\ref{#1}}
\newcommand{\tabref}[1]{{Table}~\ref{#1}}

\def\bb0{{\mathbb{0}}}

\def\ba{{\mathbf{a}}}
\def\bb{{\mathbf{b}}}

\def\bff{{\mathbf{f}}}
\def\bg{{\mathbf{g}}}
\def\bh{{\mathbf{h}}}

\def\bn{{\mathbf{n}}}

\def\b0{{\mathbf{0}}}

\def\bH{{\mathbf{H}}}

\def\bX{{\mathbf{X}}}

\def\sf0{{\mathsf{0}}}

\usepackage{epstopdf}
\usepackage{enumerate}
\usepackage{algorithm}
\usepackage{amsmath}
\usepackage{float}
\usepackage{color}
\usepackage{makeidx}
\usepackage{bm}
\usepackage{cleveref}
\usepackage{url}
\usepackage{steinmetz}
\usepackage{varwidth}% http://ctan.org/pkg/varwidth
\usepackage{soul}
\graphicspath{{figs/}}
\usepackage{balance}

\newcommand{\sref}[1]{{Section}~\ref{#1}}

\DeclareMathOperator*{\argmax}{arg\,max}
\DeclareMathOperator*{\argmin}{arg\,min}

\newcommand{\abs}[1]{\lvert#1\rvert}

\begin{document}

\title{Radar Aided 6G Beam Prediction: Deep Learning Algorithms and Real-World Demonstration}

\author{Umut Demirhan and  Ahmed~Alkhateeb
	\thanks{The authors are with the School of Electrical, Computer and Energy Engineering, Arizona State University, Tempe, AZ, 85281 USA (Email: udemirhan, alkhateeb@asu.edu).}}

\maketitle

\begin{abstract}
	
	This paper presents the first machine learning based real-world demonstration for radar-aided beam prediction in a practical vehicular communication scenario. Leveraging radar sensory data at the communication terminals provides  important awareness about the transmitter/receiver locations and the surrounding environment. This awareness could be utilized to reduce or even  eliminate the beam training overhead in millimeter wave (mmWave) and sub-terahertz (THz) MIMO communication systems, which enables a wide range of highly-mobile low-latency applications. In this paper, we develop deep learning based radar-aided beam prediction approaches for mmWave/sub-THz systems. 	The developed solutions leverage domain knowledge for radar signal processing to extract the relevant features fed to the learning models. This optimizes their performance, complexity, and inference time. The proposed radar-aided beam prediction solutions are evaluated using the large-scale real-world dataset DeepSense 6G, which comprises co-existing mmWave beam training and radar measurements.  In addition to completely eliminating the radar/communication calibration overhead, the experimental results showed that the proposed algorithms are able to achieve around $90\%$ top-5 beam prediction accuracy while saving $93\%$ of the beam training overhead. This highlights a promising direction for addressing the beam management overhead challenges in mmWave/THz communication systems.

\end{abstract}

\section{Introduction}
Millimeter wave (mmWave) and terahertz (THz)  communications systems rely on the beamforming gains of the narrow beams to achieve sufficient receive signal power. Finding the best narrow beam (or beam pair), however, requires high beam training overhead, which makes it hard for these systems to support highly mobile applications such as vehicular, drone, or augmented/virtual reality communications \cite{Alkhateeb2018a}. One important observation here is that the beam selection problem highly relies on the transmitter/receiver locations and the geometry/characteristics of the surrounding environment. This means that acquiring some awareness about the surrounding environment and the transmitter/receiver locations could potentially help the mmWave beam selection problem. An efficient way to acquire this awareness is by using the low-cost radar sensors such as those initially designed for radar applications \cite{Ginsburg} or by leveraging joint communication-radar systems \cite{Kumari_2018,Taha2021}. \textbf{With this motivation, this paper investigates the potential of leveraging radar sensory data to guide the beam selection problem and provides the first machine learning based real-world demonstration for radar-aided beam prediction in a practical vehicular communication scenario.}

\begin{figure}[!t]
	\centering
	\includegraphics[width=1\linewidth]{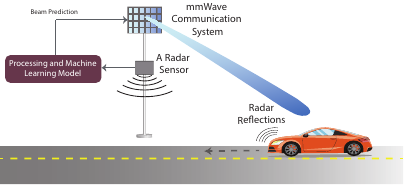}
	\caption{The system model where the radar information at the basestation is leveraged to select the  beam that serves the mobile user. }
	\label{fig:systemmodel}
\end{figure}

Leveraging sensory data to guide the mmWave beam selection problem has gained increasing interest in the last few years \cite{Alrabeiah2020,Ali2018, charan2021visionposition,Va_position, Alrabeiah2020a, ali2019millimeter,Yuan_radar}. In \cite{Alrabeiah2020,Ali2018}, the authors proposed to leverage the sub-6GHz channels that are relatively easier to acquire to guide the beam selection problem. Acquiring sub-6GHz channels, however, still requires allocating wireless communication resources and probably additional control signaling.  In \cite{charan2021visionposition,Va_position}, the position information of the user was leveraged by the base station to select the mmWave beam. The position information, though, may not be sufficient to accurately determine the best beam, which is also a function of the surrounding environment, especially in the non-line-of-sight scenarios. Further,  acquiring accurate enough position information to adjust the narrow beams (i) may require expensive positioning systems  at the user for the outdoor scenarios, and (ii) is hard to achieve for indoor communication.  This motivated leveraging other data modalities for beam selection such as vision \cite{charan2021visionposition,Alrabeiah2020a}, which could be acquired at low-cost and without consuming any wireless communication/control resources, or radar data \cite{ali2019millimeter, Yuan_radar} which may operate at a different band than that used by the mmWave communication system. The prior work on using radar for beam management, however, relied mainly on classical calibration techniques for the radar and communication systems, which could be expensive and hard to implement in reality. The prior work was also evaluated only using  computer simulations and relatively simple scenarios that are different from real-world deployments and practical hardware imperfections.

In this work, we develop  machine learning based algorithms for radar-aided mmWave beam prediction and demonstrate their performance using a real-world dataset  in a realistic vehicular communication scenario. The main contributions of the paper can be summarized as follows: (i) We formulate the radar-aided beam prediction problem considering practical radar and communication models, (ii) we then develop efficient machine learning algorithms that leverage classical signal preprocessing approaches for extracting the relevant features such as range-velocity, range-angle, and range-velocity-angle maps, (iii) leveraging the large-scale real-world dataset, DeepSense 6G \cite{DeepSense} that comprises co-existing mmWave beam training and radar measurements,  we evaluate and demonstrate the performance of the proposed radar-aided beam prediction approaches in a realistic vehicular communication scenario. We also draw important insights about the trade-offs of the various algorithms in terms of beam prediction accuracy, processing time, inference latency, and complexity overhead.

\section{System Model} \label{sec:systemmodel}
The considered  system in this paper consists of a base station and a mobile user. The base station employs two main components: (i) A mmWave communication terminal equipped with a phased array that is used to communicate with the mobile user, and (ii) an FMCW radar that is leveraged to aid the selection of the mmWave communication beam. The system model is illustrated in \figref{fig:systemmodel}. In the next two subsections, we briefly describe the system and signal models of the communication and radar components. For ease of exposition, we summarize the adopted notation in \tabref{table:notation}.

\begin{table}[!t]
	\centering
	\caption{Notation adopted in the paper}
	\label{table:notation}
	\renewcommand{\arraystretch}{1.2}
	\resizebox{\columnwidth}{!}{
		\begin{tabular}{|c|c|c|c|}
			\hline
			\textbf{Notation} & \textbf{Description}        & \textbf{Notation} & \textbf{Description}  \\ \hline
			$M_r$               & \# of radar RX antennas     & $T_c$             & Chirp duration        \\ \hline
			$S$               & \# of samples per chirp     & $T_f$             & Frame duration        \\ \hline
			$A$               & \# of chirps per frame      & $\bX_l$           & Radar data of frame $l$ \\ \hline
			$L$               & \# of frames                & $\bH_{\mathrm{RC}}$     & Radar cube            \\ \hline
			$N$               & Codebook size               & $\bH_{\mathrm{RA}}$               & Range-Angle maps      \\ \hline
			$\mu$             & Chirp slope                 & $\bH_{\mathrm{RV}}$               & Range-Velocity maps   \\ \hline
		\end{tabular}
	}
\end{table}

\subsection{Radar Model}
In our system, the base station adopts an FMCW radar. The objective of this radar is to provide observations of the environment. The FMCW radar achieves this objective by transmitting chirp signals whose frequency changes continuously with time. More formally, the FMCW radar transmits a linear chirp signal starting at an initial frequency $f_c$ and linearly ramping up to $f_c + \mu t$, given by
\begin{equation}
s^\textrm{tx}_\textrm{chirp}(t) = 
\begin{cases}
	\sin( 2\pi [f_c \, t + \frac{\mu}{2} \, t^2]) & \text{if }  0\leq t \leq T_c \\
	0 & \text{otherwise}
\end{cases}
\end{equation}
where $\mu= B/T_c$ is the slope of the linear chirp signal with $B$ and $T_c$ representing the bandwidth and duration of the chirp.

A single radar measurement is obtained from the frame of duration $T_f$. In each frame, $A$ chirp waves are transmitted with $T_s$ waiting time between them. After the transmission of the last chirp, no other signals are transmitted until the completion of the frame. Mathematically, we can write the transmitted signal of the radar frame as
\begin{equation}
	s^\textrm{tx}_\textrm{frame}(t) = \sqrt{\mathcal{E}_t} \sum_{a=0}^{A-1} s_\textrm{chirp}(t - (T_c+T_s)\cdot a), \quad 0\leq t\leq T_f
\end{equation}
where $\sqrt{\mathcal{E}_t}$ is the transmitter gain. The given transmitted signal is reflected from the objects in the environment, and received back at the radar. 

At the receiver, the signal obtained from an antenna is passed through a quadrature mixer that combines the transmit with receive signals resulting in the in-phase and quadrature samples. After that, a low-pass filter is applied to the mixed signals. The resulting signal, referred to as intermediate frequency (IF) signal, reflects the frequency and phase difference between the transmit and receive signals. If a single object exists in the environment, then the receive IF signal of a single chirp can be written as
\begin{equation}
	s_\textrm{chirp}^\textrm{rx}(t) = \sqrt{\mathcal{E}_t \mathcal{E}_r} \exp\left(j2\pi [\mu \tau t + f_c \tau - \frac{\mu}{2} \tau^2]\right),
\end{equation}
where $\sqrt{\mathcal{E}_r}$ is the channel gain of the object which depends on the radar cross section (RCS) and the path-loss, $\tau = 2d/c$ is the round-trip delay of the reflected signal through the object with $d$ denoting the distance between the object and the radar, and $c$ representing the speed of light. 

The receive IF signal, $s_\textrm{chirp}^\textrm{rx}(t)$ is then sampled at the sampling rate of the ADC, $f_s$, producing $S$ samples for each chirp. Finally, the ADC samples from each frame are collected. For an FMCW radar with $M$ receive antennas, each having the described RF receive chain, the resulting measurements (raw-data) of one frame can be denoted by $\bX \in \mathbb{C}^{M_r\times S\times A}$. In the following subsection, we describe the communication model.

\begin{figure*}[!t]
	\centering
	\includegraphics[width=.7\linewidth]{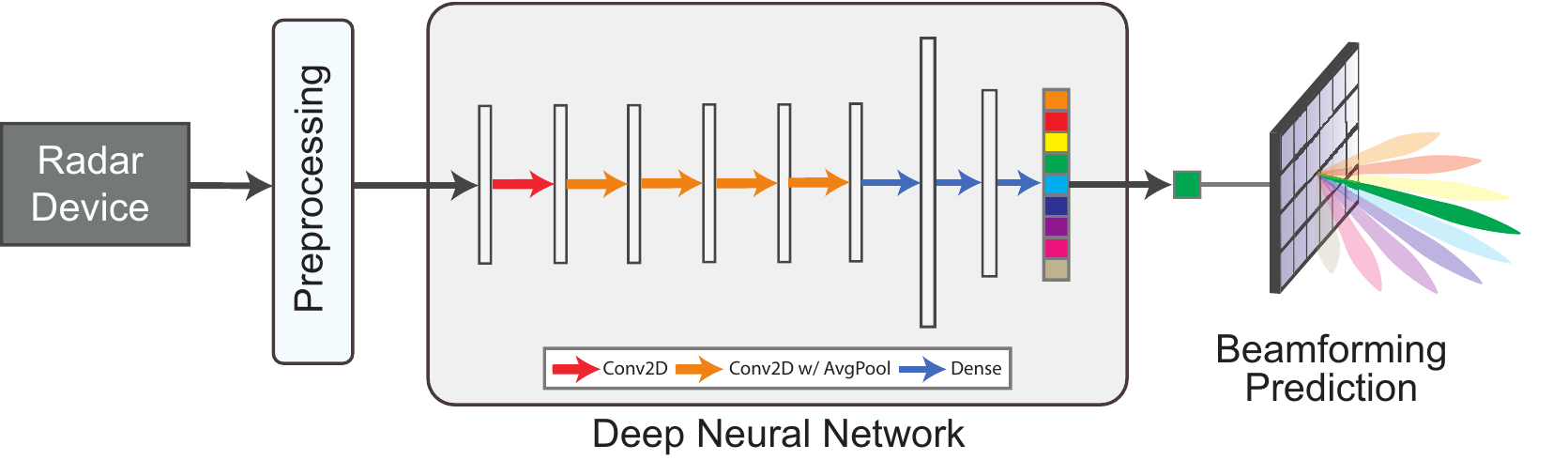}
	\caption{The figure illustrates the proposed approach where the radar observations are preprocessed to extract the useful features. These features are then fed to the deep neural network model, which returns predicts the beamforming vector that should be used at the basestation.}
	\label{fig:solutionapproach}
\end{figure*}

\subsection{Communication Model}
The considered base station employs a mmWave transceiver with $M_\mathrm{A}$ antennas and use it to communicate with a single-antenna mobile user. Adopting a narrowband channel model, the channel between the user and the base station can be expressed as
\begin{equation}
    {\bh} = \sum_{p=0}^{P-1} \alpha_p \ba(\phi_{p}, \theta_{p})
\end{equation}
where $\alpha_p$ denotes the complex gain and $\phi_{p}$, $\theta_{p}$ represent transmit azimuth and elevation angles of the $p$-th path at the base station. In the downlink, the base station transmits the data symbol $s_d$ to the user via the beamforming vector $\bff \in \mathbb{C}^{M_c}$. The receive signal at the user can be written as
\begin{equation} \label{eqn:communicationchannel}
    y = \sqrt{\mathcal{E}_c} {\bh}^H \bff s_d + n
\end{equation}
where $n \sim \mathcal{CN}(0, \sigma^2)$ is the additive white Gaussian noise and $\sqrt{\mathcal{E}_c}$ is the transmitter gain of the basestation. For the selection of the beamforming vectors, we define the beamforming codebook of $N$ vectors by $\bm{\mathcal{F}}$, where the $n$-th beamforming vector is denoted by $\bff_n \in \bm{\mathcal{F}},\ \forall n \in \{0, \ldots, N-1\} $ \cite{Zhang2021,Li_hybrid_2019}. Hence, $\bff$ in \eqref{eqn:communicationchannel} is restricted to the beams in the codebook. With this model, the index of the optimal beam, $n^\star$, can be obtained by the SNR maximization problem, i.e.,
\begin{align}
    n^\star = \underset{n}{\textrm{argmax}} \ \abs{\bg^H \bff_n}^2 \quad \textrm{s.t.} \quad \bff_n \in \mathcal{\bm{\mathcal{F}}}
\end{align}
where the optimal solution can be obtained by an exhaustive search over the possible beamforming vectors.

\section{Machine Learning for Radar Aided Beam Prediction: Problem Formulation} \label{sec:ML+BF}
In this section, we formally define the radar based beamforming problem, building upon the described system model in \sref{sec:systemmodel}. Then, we present the key idea of the proposed solution. Finally, we define the beam prediction machine learning task.

\subsection{Problem Definition}
In this paper, we seek to leverage the radar measurements $\bX$ in determining the optimal communication beamforming vector $\bff_{n^\star}$. First, let us introduce the subscript $l$ to indicate the $l$-th radar frame. The radar measurements during this frame will then be denoted as $\bX_l$. Further, we add this subscript $l$ to the beamforming index and the beamforming vector used in this $l$-th frame, to be $n_l$, $\bff_{n_l}$.  If a single-user exists in the line-of-sight (LOS) of the base station, then the radar measurements could potentially include useful information $\bX_l$ about its position/orientation with respect to the base station. This position/orientation information could be leveraged to guide the optimal beam selection. To formulate that, we define the mapping function $\Psi_\mathbf{\Theta}$ to capture the mapping from the radar observations to the optimal beamforming index, given by
\begin{equation}
    \Psi_{\bf \Theta}: \{\bX_l\} \rightarrow \{n^\star_l\}
\end{equation}  
Our objective is then to design the mapping function $\Psi_{\bf \Theta}$ to be able to map the radar measurements to the optimal beam index $n^\star$. Towards this objective, we investigate the possible designs of the mapping function, and learn the set of parameters $\bf \Theta$. Mathematically, we can express this objective by the following optimization problem that aims at finding the mapping function and the optimal set of parameters $\bf \Theta$, that maximizes the accuracy in predicting the optimal beam: 
\begin{equation} \label{eqn:optimization}
	\Psi^\star_{\bf\Theta^\star} = \argmax_{\Psi_\Theta} \frac{1}{L} \sum_{l=1}^L \mathbf{1}_{\{n_l^* = \Psi_{\bf \Theta}(\bX_l) \}}
\end{equation}
where $\mathbf{1}_{E}$ is the indicator function of the event $E$, i.e., $\mathbf{1}_{E} = 1$ if $E$ occurs, and $\mathbf{1}_{E} = 0$ otherwise. Next, we present the motivation of our machine learning based approach for addressing this radar-aided beam prediction problem.

\subsection{Motivation for Machine Learning}

In this paper, we propose to leverage machine learning to optimize the mapping from the radar measurements to the optimal communication beamforming vectors. The motivation of the proposed approach is based on the following main observations:
\begin{itemize}
	\item The FMCW radars are designed to collect specific measurements of the environment that are useful for the automotive applications. These measurements are generally different than the communication channel state information that we typically need for adjusting the communication beams. Therefore, a straightforward mapping between the radar and communication channels is non-trivial.
	\item Using classical approaches such as lookup tables to directly map the radar object detection and positioning information to the best beam may not be efficient. The reason goes back to the imperfections of the practical radar systems, that impact the accuracy of the detection and localization performance. Moreover, the other elements (e.g., pedestrians, bikers) in the real-word environments present themselves as a challenge in the detection and localization of the target users. Further, the beam patterns of the practical systems does not cover the field of view in an ideal way, and selecting the best beam may require additional refinement in the angle-beam mapping. Therefore, the practical features of the system and deployment scenarios degrade the performance of potential classical solutions, even in the more ideal LOS scenarios. We will elaborate more on this point in \sref{sec:results}.
	\item With the recent advances in artificial intelligence, the machine learning models became prevalent in complex mapping and recognition problems. It is mainly thanks to their high capability in extracting the inherent information with significant success rates. Particularly with the FMCW radars, the machine learning based solutions has shown significant improvements for the object, vehicle and pedestrian detection and classification problems \cite{lin2016design, zhang2020object}. Moreover, it enabled more advanced mappings such as gesture detection with radars \cite{dekker2017gesture}.
\end{itemize}
With this motivation, we propose to leverage machine learning, and in particular deep learning models, to learn the mapping from the radar measurements to the optimal beamforming vectors. Our solutions will integrate this machine learning models with domain-knowledge based radar preprocessing techniques to reduce the complexity of the learning problem in realistic environments. The general flow of the proposed machine learning and radar-aided beam prediction approach is illustrated in \figref{fig:solutionapproach}. In the next section, we will provide a detailed description of the proposed solution.

\subsection{Machine Learning Task: Radar aided Beam Prediction}
We define the machine learning task as follows: Given the $l$-th radar observation matrix (raw data) $\bX_l$, the objective is to design a machine learning model that returns the index of the optimal beam. In other words, the machine learning model aims to return the index of the beam providing the most gain in the beamforming codebook. In a more general sense, the top-$K$ predictions can be utilized. In this case, the ordered set of the $K$ most likely beam indices are returned by the model. Mathematically, for an observation $\bX_l$, the beam prediction task returns the ordered set of the indices of the most promising $K$ beams $\hat{\bn}_l^\star = \{\hat{n}_{l,1}^\star, \ldots, \hat{n}_{l,K}^\star \}$ from a codebook of $N$ vectors, i.e., $n_{l,k_1}^\star, n_{l,k_2}^\star \in \{1, \ldots, N\}$ $\forall k_1, k_2\in\{1, \ldots, K\}$ and $k_1 \neq k_2$. With this notation, the beam prediction task is a multi-class classification problem given by
\begin{equation}
	\min \frac{1}{L} \sum_{l=1}^L \mathcal{L}\left(\hat{\bn}_{l}^\star, n_l^\star\right),
\end{equation}
where $\mathcal{L(., .)}$ denotes the loss function. Regardless of the loss function, different evaluation metrics for the top-$K$ predictions can be provided. For instance, the top-$K$ accuracy of the model can be written as
\begin{equation}
	\text{Top-}K\text{ Accuracy} = \frac{1}{L} \sum_{l=1}^L \sum_{k=1}^K \mathbf{1}_{\{\hat{n}_{l,k}^\star = n_l^\star\}}.
\end{equation}
The defined machine learning task and evaluation metric will be utilized in the following sections. Next, we present the proposed machine learning aided solution.

\section{Machine Learning for Radar Aided Beam Prediction: Proposed Framework} \label{sec:solution}
In this section, we present our deep learning and radar aided beam prediction approach. Our solution integrates radar preprocessing and deep neural networks. This targets reducing the complexity of the learning task and enables efficient training with reasonable dataset sizes. To formalize our approach, we first decompose the radar to beam mapping function into three components: (i) The preprocessing function $\Psi^{\mathrm{P}} (.)$, (ii) the neural network function of the parameters $\bf \Theta$, $\Psi^{\mathrm{N}}_{\bf{\Theta}} (.)$, and (iii) the evaluation function $\Psi^{\mathrm{E}}(.)$. Then, we can write the radar to beam mapping function as
\begin{equation}
\Psi_{\bf{\Theta}}(\bX) = \Psi^{\mathrm{E}} (\Psi^{\mathrm{N}}_{\bf{\Theta}} (\Psi^{\mathrm{P}} (\bX))).
\end{equation}
With the decomposition, we can define our solution in terms of the preprocessing, neural networks, and evaluation functions. In the following, we present our approach via the subsections of each function. First, we describe the proposed preprocessing approach.

\begin{figure*}[!t]
	\centering
	\includegraphics[width=0.9\linewidth]{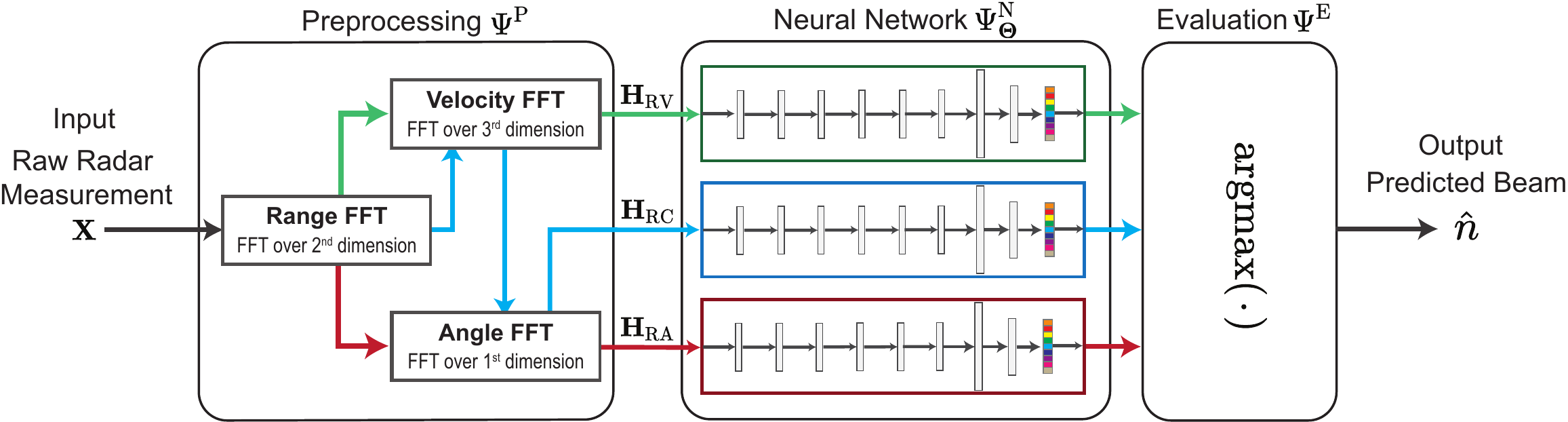}
	\caption{The figure illustrates the radar processing procedures for the three proposed approaches. The results of these processing procedures, ${\bH}_{\mathrm{RC}}, {\bH}_{\mathrm{RA}}, {\bH}_{\mathrm{RV}}$ are then presented  to the neural networks.}
	\label{fig:radarproc}
\end{figure*}

\subsubsection{Preprocessing}

In principal, given the radar raw measurements $\bX \in\mathbb{C}^{M_r\times S \times A}$, three important features that could be extracted are the range, the angles, and the velocity of the moving objects in the environment. Based on that, we propose three different preprocessing approaches, as illustrated in \figref{fig:radarproc}. Each approach leverages a certain set of these quantities. To mathematically define the preprocessing approaches, let us denote the 2D and 3D Fourier transforms by $\mathcal{F}_{\textrm{2D}(.)}$ and $\mathcal{F}_{\textrm{3D}}(.)$, respectively. Next, we describe the three preprocessing approaches. 

\textbf{Range-Angle Maps:} The first approach aims at utilizing the range and angle information. For this purpose, first, with an FFT in the direction of the time samples, referred to as the Range FFT, we obtain the chirp signal in the frequency domain. In this domain, the chirp signals are shifted proportionally to the round-trip travel duration of the signal, providing the range information. After that, a clutter removal operation can be applied to obtain cleaner images by a mean removal over the chirp samples. Then, with an another FFT in the direction of the receive antenna samples, referred to as the Angle FFT, the angular information can be obtained. An FFT of a larger size, $M_F$, can be applied with zero-padding to over-sample the angles. Finally, we can construct the final range-angle map by combining resulting range-angle information for each chirp sample. In a simplified way, the described operation can be mathematically written as 
\begin{equation}
	{\bH}_{\mathrm{RA}} = \Psi^\mathrm{P}_\mathrm{RA}(\bX) = \sum_{a=1}^A \abs{\mathcal{F}_\textrm{2D}(\bX_{:,:,a})}.
\end{equation}

\textbf{Range-Velocity Maps:} Alternatively, we consider the range-velocity maps. To construct these maps from the radar measurements, two FFTs through the time samples and chirp samples are applied. Similarly to the previous approach, first, the Range FFT is utilized. Differently, the second FFT is applied through the chirp samples, referred to as the Velocity FFT. It simply returns the phase shift over the consecutive chirp samples. This phase shift is caused by the Doppler shift, and it contains the velocity information. Finally, again by combining the range-velocity information of the different receive antenna samples, we obtain the final range-velocity map. This operation can be written as 
\begin{equation}
	{\bH}_{\mathrm{RV}} = \Psi^\mathrm{P}_\mathrm{RV}(\bX) = \sum_{m=1}^M \abs{\mathcal{F}_\textrm{2D}(\bX_{m,:,:})}.
\end{equation}

\textbf{Radar Cube:} The previous approaches combine the angle or velocity dimensions, reducing the information to a 2D map. Without a dimensionality reduction, we apply the range, velocity, and angle FTTs, and obtain the radar cube. The resulting radar cube contains all the information of the range, velocity, and angle of the targets. It can be considered as the stack of range-angle maps of each velocity value. The operation can be mathematically described as
\begin{equation}
	{\bH}_{\mathrm{RC}} = \Psi^\mathrm{P}_\mathrm{RC}(\bX) = \abs{\mathcal{F}_{\textrm{3D}}(\bX)}.
\end{equation}

\begin{table}[!t]
	\caption{Complexity and Memory Requirements of Different Inputs}
	\label{table:preprocessing}
	\centering
	\begin{tabular}{|c|c|c|}
		\hline
		\textbf{Network Input} & \textbf{Preprocessing Complexity}                & \textbf{Input Size} \\ \hline
		${\bH}_{\mathrm{RC}}$          & $\mathcal{O}(M_r S A (\log S+\log A + \log M_r))$  & $M_r S A$           \\ \hline
		${\bH}_{\mathrm{RA}}$                  & $\mathcal{O}(M_r S A \log S + M_F S A \log M_F )$	      & $M_{F} S$             \\ \hline
		${\bH}_{\mathrm{RV}}$                  & $\mathcal{O}(M_r S A (\log S + \log A))$           & $S A$               \\ \hline
	\end{tabular}
\end{table}

After the alternative modalities of the radar information are extracted, the data is standardized and fed into the neural networks. The described radar processing approaches bring different preprocessing complexity and input size. In particular, while the radar cube requires a 3D FFT presenting the most detailed information, it suffers from the high number of dimensions. In contrast, the range-angle and range-velocity images only require 2D FFTs and provide smaller input sizes. The further evaluation of the complexity is carried out in \sref{sec:results}. 
Next, we present the deep neural networks adopted for each modality of the data.

\subsubsection{Neural Network Modeling}

For the neural networks, to keep the complexity of the approach low, we rely on a comparably simple deep learning model with a design with convolutional and fully-connected (FC) layers. Specifically, the deep neural networks (DNNs) comprise 8 total layers. The first five layers are the convolutional layers with the rectified linear unit (ReLU) activation functions. In addition, the average pooling is applied after the activation of the convolutional layers to decrease the size of the data. Finally, the output of the fifth convolutional layer is connected to a set of three FC layers, providing $N$ outputs. The each entry of the output indicates a beam.

As the proposed inputs of the neural networks are of different size and dimensions, the same network cannot be applied to the all types of the inputs. Therefore, for different modalities of the radar data, the input, output and kernel size of the DNN layers are adjusted to keep the network size reasonable and similar while providing a comparably good performance. Specifically, we adjust the networks for our dataset, which will be described in \sref{sec:datacollection}. In this dataset, the system parameters are given by $S=256$, $A=128$, $M_r=4$ and $N=64$. With these parameters and $M_F \in \{4, 64\}$, the designed DNN architectures are summarized in Table \ref{table:neuralnetworks} \footnote{The additional $M_F=64$ point angle FFT is only applied for the range-angle maps. For the radar-cube, only $M_F=M_r=4$ point angle FFT is applied to keep the input size of the different data modalities reasonably similar.}.

\begin{table}[!t]
	\centering
	\caption{Deep Neural Network Architectures for Different Input Types}
	\label{table:neuralnetworks}
	\renewcommand{\arraystretch}{1.4}
	\resizebox{\columnwidth}{!}{
		\begin{tabular}{|c|c|c|c|c|}
			\hline
			\textbf{NN Layers} & \textbf{Radar Cube ($\bH_{\mathrm{RC}}$)}           & \textbf{Range-Velocity ($\bH_{\mathrm{RV}}$)}     & \textbf{Range-Angle-64 ($\bH_{\mathrm{RA}}$)} & \textbf{Range-Angle-4 ($\bH_{\mathrm{RA}}$)}        \\ \hline
			\textbf{Input}     & $4 \times 256 \times 128$   & $1 \times256 \times 128$    & $1 \times 256 \times 64$ & $1 \times 256 \times 4$     \\ \hline
			\textbf{CNN-1}     & \multicolumn{4}{c|}{Output Channels: 8, Kernel: (3, 3), Activation: ReLU}  \\ 
			\hline
			\textbf{CNN-2}     & \multicolumn{4}{c|}{Output Channels: 16, Kernel: (3, 3), Activation: ReLU}  \\ 
			\hline
			\textbf{AvgPool-1} & \multicolumn{2}{c|}{Kernel: (2, 1)}     & \multicolumn{2}{c|}{N/A}           \\ 
			\hline
			\textbf{CNN-3}     & \multicolumn{4}{c|}{Output Channels: 8, Kernel: (3, 3), Activation: ReLU}  \\ 
			\hline
			\textbf{AvgPool-2} & \multicolumn{3}{c|}{Kernel: (2, 2)} &    Kernel: (2, 1)       \\ \hline
			\textbf{CNN-4}     & \multicolumn{4}{c|}{Output Channels: 4, Kernel: (3, 3), Activation: ReLU}  \\ 
			\hline
			\textbf{AvgPool-3} & \multicolumn{3}{c|}{Kernel: (2, 2)} &    Kernel: (2, 1)    \\ \hline
			\textbf{CNN-5}     & \multicolumn{4}{c|}{Output Channels: 2, Kernel: (3, 3), Activation: ReLU}  \\ 
			\hline
			\textbf{AvgPool-4} & \multicolumn{3}{c|}{Kernel: (2, 2)} &    Kernel: (2, 1)       \\ \hline
			\textbf{FC-1}     & \multicolumn{4}{c|}{Input Size: 512, Output Size: 256, Activation: ReLU}  \\ 
			\hline
			\textbf{FC-2}     & \multicolumn{4}{c|}{Input Size: 256, Output Size: 128, Activation: ReLU}  \\ 
			\hline
			\textbf{FC-3}     & \multicolumn{4}{c|}{Input Size: 128, Output Size: 64}  \\ 
			\hline 
		\end{tabular}	
	}
\end{table}

\textbf{Neural Network Objective:} To train the neural networks with the aim of finding the optimal parameters $\bf \Theta^\star$, we can write the following optimization problem that aims at minimizing the loss between the output of the network and the optimal beam values, $n_l^\star \in \{0, \ldots, N-1 \}$:
\begin{equation} \label{eqn:nnobjective}
	{\bf \Theta^\star} = \argmin_{\bf \Theta} \frac{1}{L} \sum_{l=1}^L \mathcal{L}\left(\Psi^{\mathrm{N}}_{\bf{\Theta}}\big(\Psi^\mathrm{P}(\bX_l)\big), n_l^\star\right)
\end{equation}
where $\mathcal{L(., .)}$ denotes the loss function, which should be selected based on the problem type. As our problem is a multi-class classification problem, we utilize the cross-entropy loss given by
\begin{equation}
	\mathcal{L}(\hat{\bb}, \bb) = -\frac{1}{N} \sum_{n=0}^{N-1} b_n \log(\hat{b}_n)
\end{equation}
where $\bb = [b_0, \ldots, b_{N-1}]$ is the one-hot encoded vector of the optimal beam $n^\star_l$ and $b_n=\Psi^{\mathrm{N}}_{\bf{\Theta}} (\Psi^{\mathrm{P}} (\bX_l))$ is the output of the neural network. To clarify, the elements of the one-hot encoded vector $\bb$ is defined by
\begin{equation}
	b_n =
	\begin{cases}
		1, &\textrm{if } n=n^*, \\ 
		0, &\textrm{otherwise}.
	\end{cases}
\end{equation} 
Using the defined loss function, the neural network can be trained with back-propagation through the layers. We note that by the construction of proposed neural networks, the DNN models return the soft information, $\hat{\bb} \in \mathbb{R}^N$, which needs to be converted to the beam indices. Next, we describe the evaluation function of our approach.

\begin{figure*}[!t]
	\centering
	\subfigure[The basestation antenna array and radar device on the setup]{
		\includegraphics[width=.43\linewidth]{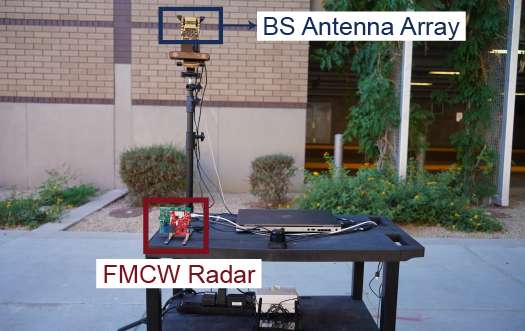}
		\label{fig:setup_im_a}
	}
	\subfigure[The system setup with a car on sight]{
		\includegraphics[width=.43\linewidth]{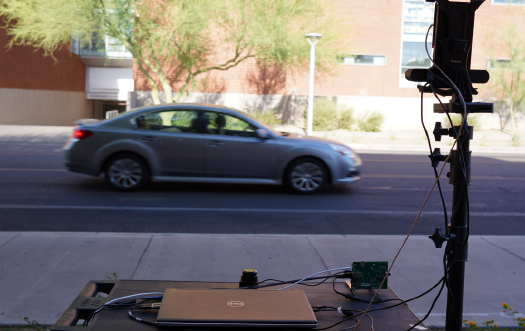}
		\label{fig:setup_im_b}
	}
	\caption{The figure illustrates the data collection setup which consists of a base station unit and a mobile unit (vehicle). The base station is equipped with a  mmWave FMCW radar and a mmWave antenna array. The car is equipped with an omni-directional mmWave transmiter. Figure (a) shows a front-view and figure (b) shows a back-view with a mobile user (vehicle) on sight.   %In the dataset, the car always moves from left-to-right on the closer lane, as shown in the picture (b).
	}
	\label{fig:datasetup}
\end{figure*}

\subsubsection{Evaluation}
To evaluate the output of the neural network in terms of the objective function in \eqref{eqn:optimization}, we need to select a single beam from the soft output of the neural network. For this purpose, the maximum of the neural network output can be selected as the prediction of the optimal beamforming vector. This can be mathematically provided by setting $\Psi^\mathrm{E}(.) = \argmax(.)$, completing our solution. In the following section, we describe our dataset adopted in the training and evaluation of the solutions.

\section{Real-World Dataset} \label{sec:datacollection}

\begin{figure*}[!t]
	\centering
	\includegraphics[width=1\linewidth]{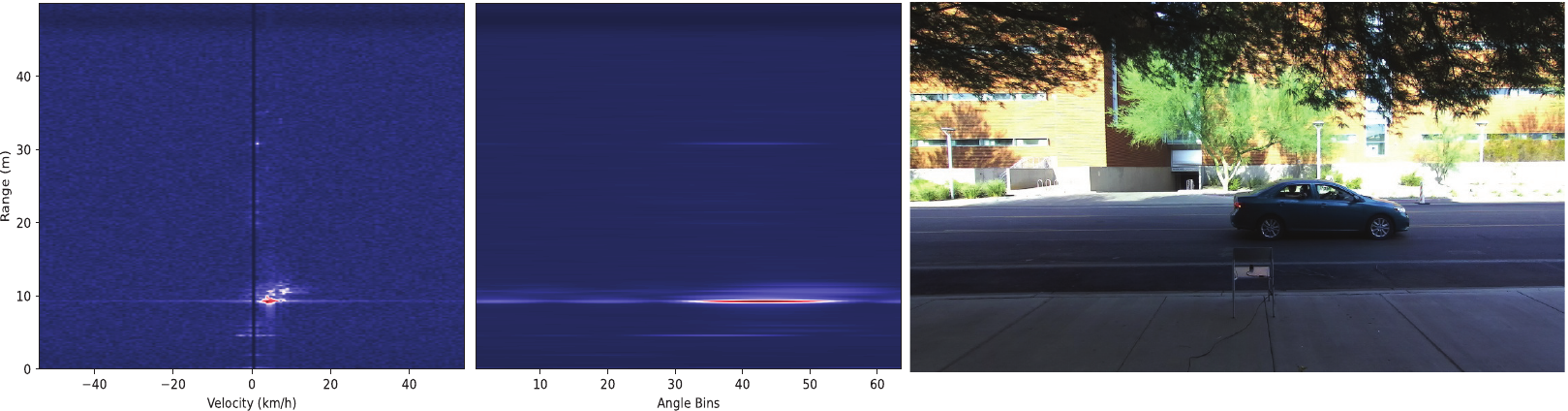}
	\caption{A sample from the dataset is shown with the current environment image (right) and the corresponding range-angle (middle) and range-velocity (left) images. The car is on the right part of the camera image, moving away from the vertical angle of the radar device. The range-angle image shows the position at approximately 9m distance on the right hand side, while the range-velocity image indicates the increasing relative velocity and range. }
	\label{fig:sample_image}
\end{figure*}

To accurately evaluate the performance of the proposed machine learning based radar-aided beam prediction approaches in a realistic environment, we built a real-world dataset with radar and wireless measurements. In this section, we describe our testbed and present the dataset collection scenario.

\subsection{Data Collection Testbed} \label{subsec:testbed}

We adopt the Testbed-1 of the DeepSense 6G dataset\cite{DeepSense}. The testbed comprises two units: The stationary unit (Unit 1), and the mobile unit (Unit 2). Among other sensors, unit 1 employs an FMCW radar (AWR2243BOOST) which has 3 transmit and 4 receive antennas, and a mmWave receiver at $60$ GHz which adopts a uniform linear array (ULA) with $M_\mathrm{A}=16$ elements. The unit 2 utilizes a $60$ GHz quasi-omni antenna, acts as a transmitter and is always oriented towards the receiver antenna of unit 1. The setup of unit 1 is shown in \figref{fig:setup_im_a}, where the receiver antenna array and FMCW radar board are placed at a close proximity.

The phased array of unit 1 utilizes an over-sampled beamforming codebook of $N=64$ vectors, which are designed to cover the field of view. It captures the receive power by applying the beamforming codebook elements as a combiner. The combiner providing the most power is taken as the optimal beamforming vector. For the radar, we only activated one of the TX antennas, while the data from $M_r=4$ RX antennas were captured. We adopted a set of radar parameters based on the TIs short range radar (SRR) example, given by $B=750$ MHz, $\mu=15$ MHz/us, $A=128$ chirps/frame, $S=256$ samples/chirp. These settings provide the maximum range of $45$m and the maximum velocity of $56$ km/h, which are well-fit for the scenario illustrated in \figref{fig:setup_im_b}. For further details, please refer to the data collection testbed description in \cite{DeepSense}. Next, we present the dataset and collection scenario.

\subsection{Development Dataset}

For the evaluation, we used the testbed described in \sref{subsec:testbed} and adopted  Scenario $9$ of the DeepSense 6G dataset \cite{DeepSense}. In this scenario, a passenger at the back seat of the car holds the transmitter. As shown in \figref{fig:sample_image}, the car passes by the stationary unit (Unit 1) which collects the radar and beam training measurements. During the data collection, the road was actively used by the other cars, pedestrians and bikers. Our testbed collected and saved the radar measurements and the received power at each communication beam. 

In the construction of the dataset, the beam providing the highest power is saved as the optimal beamforming vector. The data is cleaned by only keeping the samples with the target car in sight. This cleaning operation is performed manually through the inspection of the RGB images that are captured from a camera attached next to the antenna array. The data samples with the other elements (cars, pedestrians and bikers) are also kept to reflect the realistic environment. The final dataset comprises $6319$ samples, which are separated with a $70/30\%$ split for the training and testing. A sample from the dataset through the extracted range-angle and range-velocity images are shown in \figref{fig:sample_image}.

\section{Results} \label{sec:results}
In this section, we evaluate\footnote{The numerical computations are carried out on a server with an Intel Xeon Silver 4216 processor with an Nvidia RTX Titan GPU. The code is written with the standard scientific Python libraries and PyTorch.} and compare the performance of the proposed solutions. In particular, we compare the DNN based solutions adopting the radar cube, range-velocity and range-angle maps, respectively, and a simple baseline algorithm. In the evaluation, we adopt our dataset described in \sref{sec:datacollection}. These different solutions are compared in terms of their prediction accuracy, complexity/inference time, and the required dataset sizes. 

\textbf{Baseline Algorithm:} As the baseline algorithm, we adopt a look-up table mapping the given position of the maximum point in the range-angle image to the most-likely beams. The look-up table is constructed by using the training dataset. For the top-$K$ accuracy of this solution, we select $K$ different beams corresponding to the largest points in the range-angle map.

\textbf{Training and Evaluation:} For the evaluation of the neural networks, we trained the DNN models summarized in \tabref{table:neuralnetworks} using the Adam algorithm \cite{kingma2014adam} with a learning rate $0.001$, batch size $32$, and a decay factor $\gamma=0.1$ which is applied after every $10$ epochs. The networks are trained for $40$ epochs and the network parameters showing the best top-$1$ accuracy over the validation dataset is saved for evaluation. The network training operation is carried out for $5$ separate instances, and the average performance is shown in the following results. For the evaluation of the range-angle map based solutions (the baseline and deep learning solutions), we adopt two angle FFT values as mentioned in \sref{sec:solution}, i.e., $4$ and $64$ points angle FFTs. Next, we compare the beam prediction accuracy of the solutions.
 
\begin{figure}[!t]
	\centering
	\includegraphics[width=1\linewidth]{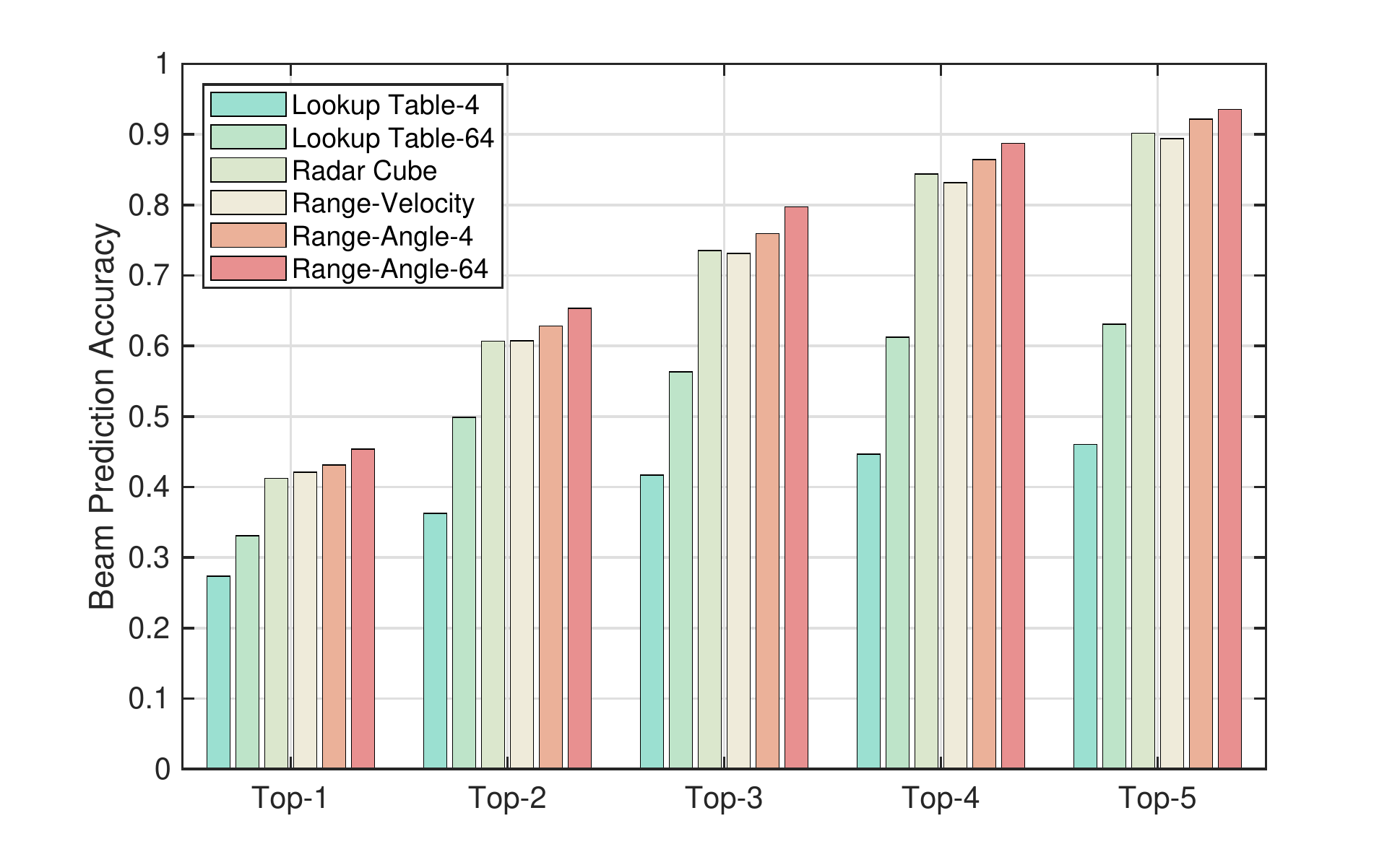}
	\caption{The top-$k$ test accuracy of the proposed approaches. Range-angle image based solution performs slightly better than the range-velocity approach, and both of them outperform the raw data based learning/prediction.}
	\label{fig:accuracy}
\end{figure}

\textbf{Beam Prediction Accuracy:} We first compare the performances of the solutions in terms of the accuracy. As shown in \figref{fig:accuracy}, the range-angle map based deep learning solutions over-perform the range-velocity and radar cube solutions. In comparison to the deep learning solutions, the baseline solutions show an inferior performance. Specifically, the baseline with $64$-point angle FFT provides $33\%$ accuracy, while the deep learning models provide at least $8\%$ better top-$1$ accuracy. This shows the robustness and applicability of the deep learning models to the real-world data. The range-velocity maps close performance to the other deep learning solutions. This is mainly due to two reasons: (i) The velocity is in the direction of the basestation, and hence, it contains an angular information. (ii) Although the right-to-left and left-to-right movement cannot be distinguished in the range-angle maps, the traffic flow allows it since the cars moving left-to-right are on the closer lane and separated by a distance. Moreover, the scenario only contains the target car moving from left-to-right as described in \sref{sec:datacollection}. The radar cubes contain the range-angle maps of different velocity values, however, it cannot perform similarly to the $4$-point range-angle solution. This is potentially due to the large size and complexity of the input and comparable simplicity of the deep learning model. It might be possible to over-perform both solutions with the radar cube adopting more complex neural networks, however, it might be computationally prohibitive to adopt the the basestation.

For top-$5$ accuracy, the baseline solution only reaches up to $63\%$ and has a smaller gain when increasing the $K$ values. In comparison, the top-$3$ and top-$5$ accuracy of the range-angle images reach up to $79.7\%$ and $93.5\%$, outperforming the baseline solution by a large margin. The other deep learning based approaches show improvements similar to the range-angle based deep learning solutions with increasing $K$ values. Based on the presented results, we conclude that the deep learning solutions show clear potential for real applications, especially with the top-$3$ and top-$5$ results. Moreover, the comparison of the range-angle solutions with different angle FFT sizes show the advantage of the maps generated with higher resolution. This is expected to be the case with the baseline solution, however, the behavior is also prevalent in the deep learning solution, which shows the potential advantages of generating maps with higher resolution. 

\begin{figure}[!t]
	\centering
	\includegraphics[width=1\linewidth]{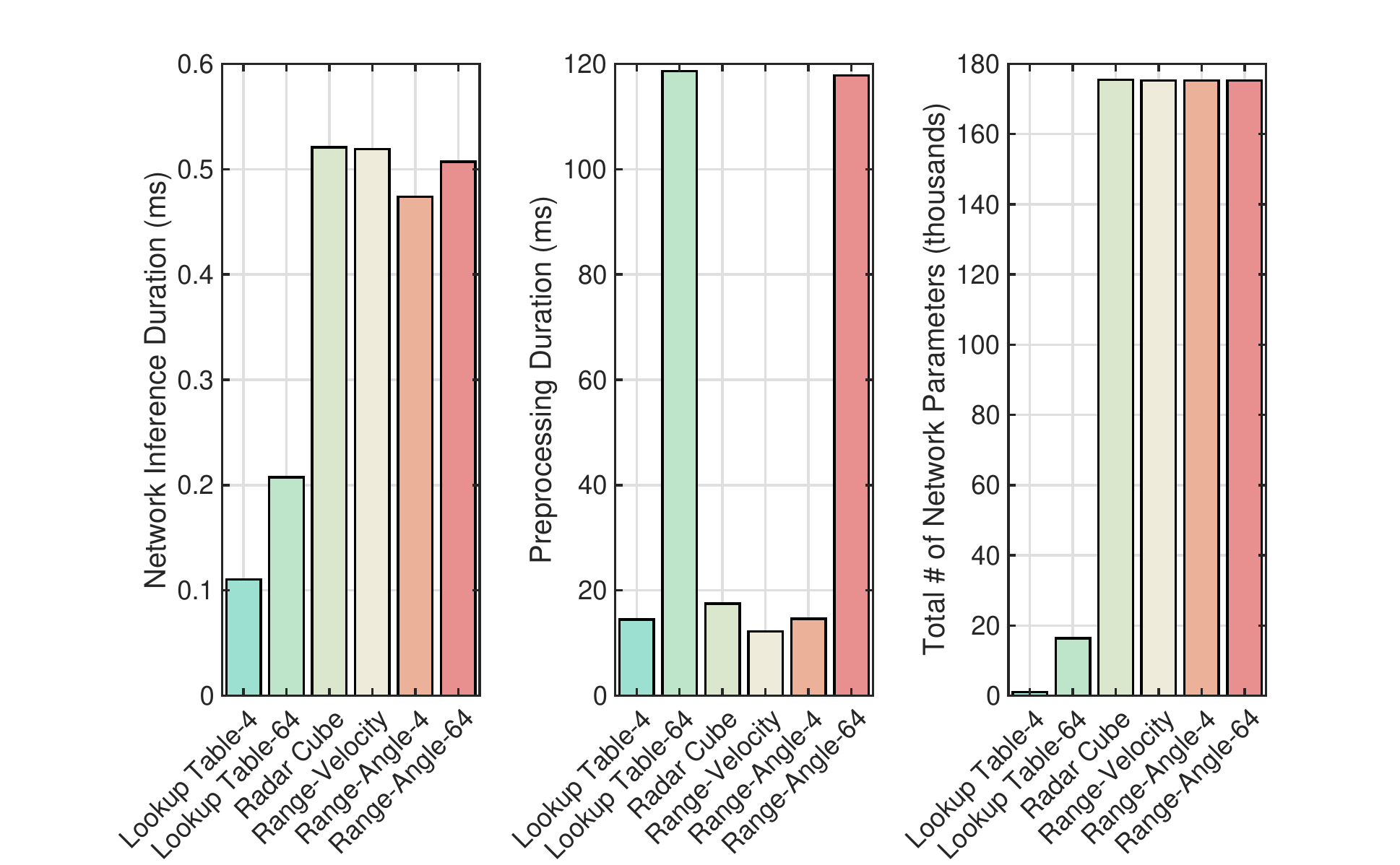}
	\caption{The complexity of the proposed radar-aided beam prediction approaches are compared in terms of the network inference time, preprocessing duration, and number of (neural network) parameters.}
	\label{fig:complexity}
\end{figure}

\textbf{Complexity}: In \figref{fig:complexity}, we compare the complexity of the DNNs in terms of the number of parameters, preprocessing and network inference durations \footnote{The DNNs are run on the GPU while the preprocessing is applied by the CPU. Therefore, they are not one-to-one comparable with each other. The durations scale based on the corresponding specialized hardware separately.}. First, in the design of the DNN models, we aimed to keep the number of parameters similar. To that end, the radar cube, range-velocity and range-angle based solutions are comprised of approximately $175$k parameters. The baseline solutions only require $1024$ and $16384$ parameters for $4$- and $64$-point angle FFTs. These values correspond to each pixel in the range-angle maps, each of which is utilized to represent the most likely beam for each point. Second, we compare the inference duration of the neural networks as shown on the left part of \figref{fig:complexity}. As one can expect, the deep learning solutions show similar inference durations since the number of parameters and network architectures are designed similarly. Third, the middle figure in \figref{fig:complexity} illustrates the radar preprocessing durations, where the larger angle FFT adopted in the range-angle maps causes a significant additional time. This present a trade-off between the beam prediction accuracy and complexity of the solution. Depending on the hardware availability, one may prefer to design a solution with higher resolution maps and better accuracy.  Without an oversampling of the angle FFT dimension, all the approaches show similar durations, presenting an advantage for the better performing solutions.

\begin{figure}[!t]
  \centering
  \includegraphics[width=1\linewidth]{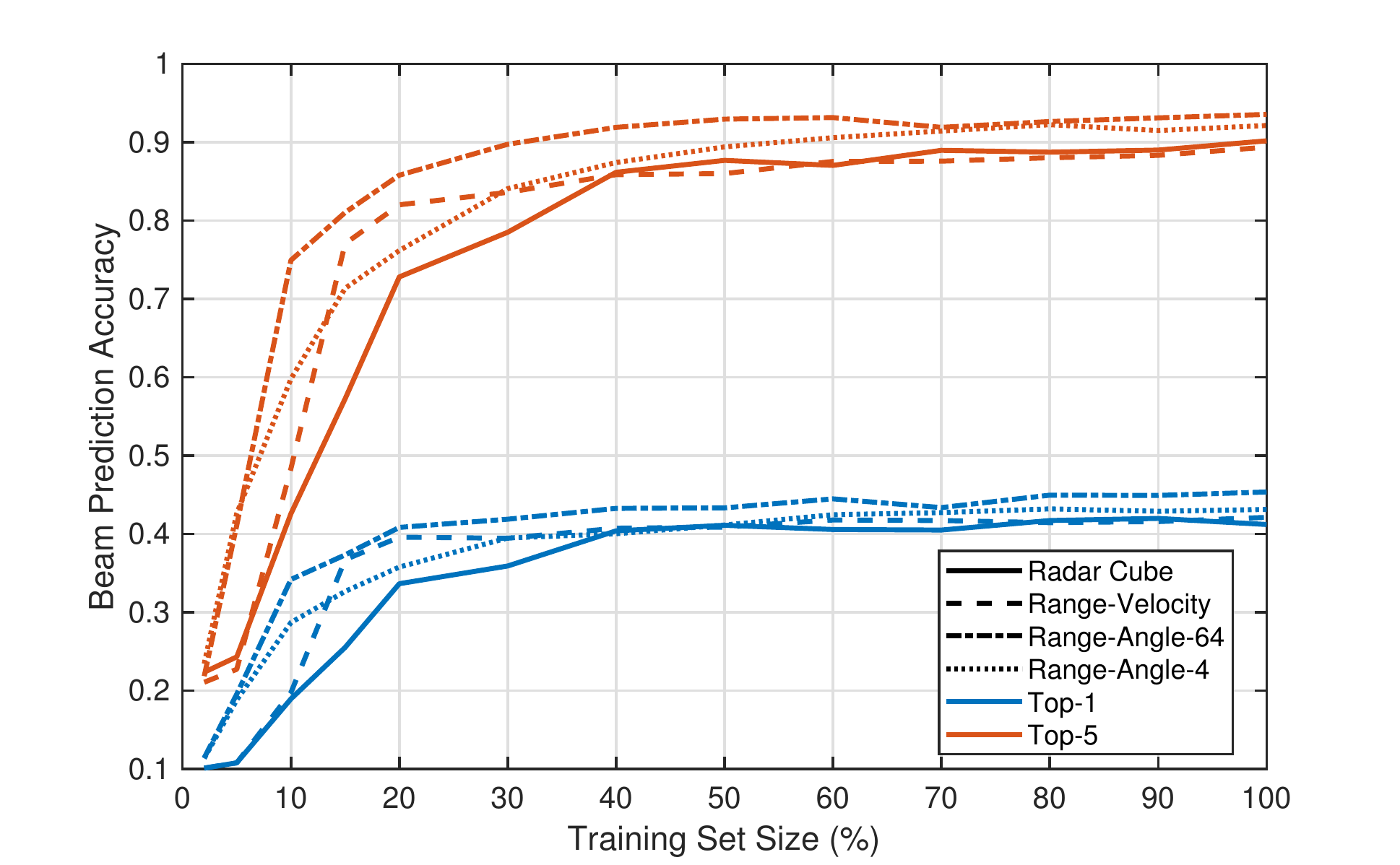}
  \caption{Top-$1$, Top-$3$ and Top-$5$ beam prediction accuracies of the proposed rada-aided beam prediction approaches when a fixed percentage of the training dataset is utilized.}
  \label{fig:datasetsize}
\end{figure}

\textbf{Impact of Dataset Size:} To draw some insights about the required dataset size, we consider different percentages of the training dataset to train the neural network models. In \figref{fig:datasetsize}, we show the average accuracy of the trained networks on the same test samples, while only a subset of the training samples are utilized. The figure shows top-$1$ and top-$5$ accuracy values. The lines of the same input for different $K$ values show similar behavior with different scaling and accuracy levels. In the figure, the accuracy of the $64$-point range-angle map based solution increases steeper than the others, reaching to a better accuracy. The radar-cube shows the slowest initial increase, possibly due to the larger input size, and eventually reaches to similar accuracy to the range-velocity solution. The performance of the $64$-point range-angle solution requires $10-20\%$ of the data for starting to saturate, while the other solutions need around $20-30\%$ of the training data. This might be due to the easier interpretability of the high-resolution range-angle maps for the beam prediction task, potentially requiring less learning and transformation. The radar-cube start to saturate particularly late, and may slightly increase with more data, indication a potential benefit from a larger dataset or more complex models. Nevertheless, the range angle/velocity based solutions generalize the problem well and can perform well with smaller datasets.

\section{Conclusion and Takeaways} \label{sec:conc}
In this paper, we developed and demonstrated, for the first time, the feasibility of radar-aided mmWave beam predictions approaches with real-world datasets. The developed solutions leveraged deep neural networks and domain-knowledge radar processing to increase the beam prediction accuracy and reduce the inference/complexity overhead. For comparisons, we have also developed a baseline  solution based on the a look-up table and the radar range-angle maps. We then evaluated the performance of the proposed solutions based on the real-world dataset DeepSense 6G, which comprises co-existing modalities including radar and mmWave beam training measurements. The main takeaways of the real-world evaluation can be summarized as follows: 
\begin{itemize}
	\item The proposed deep learning and lookup table based solutions can achieve top-$1$ beam prediction accuracies of $ 45\%$ and $33\%$, respectively, (out of a 64-beam codebook) which emphasizes the promise of leveraging radar for mmWave beam prediction and management.
	\item The top-$K$ prediction accuracy of the deep learning based solutions reach around $80\%$ for top-$3$ and $93.5\%$ for top-$5$ beams. Compared to the deep learning solutions, the classical baseline approach achieves only $56\%$ and $63\%$ top-3 and top-5 beam prediction accuracies.
	\item Among the proposed radar preprocessing approaches, the range-angle maps with different angular resolutions provide a promising trade-off between performance and complexity/inference time. For example, with $4$-point angular FFT processing, the range-angle map based solution  achieves $92\%$ top-5  beam prediction accuracy while requiring only $15$ms preprocessing and inference time. 
\end{itemize}
This real-world evaluation demonstrates the feasibility of radar-aided mmWave beam prediction and highlights its promising gains in enabling highly-mobile mmWave/sub-THz communication applications. 

\balance
% Generated by IEEEtran.bst, version: 1.14 (2015/08/26)


\begin{thebibliography}{10}
	\providecommand{\url}[1]{#1}
	\csname url@samestyle\endcsname
	\providecommand{\newblock}{\relax}
	\providecommand{\bibinfo}[2]{#2}
	\providecommand{\BIBentrySTDinterwordspacing}{\spaceskip=0pt\relax}
	\providecommand{\BIBentryALTinterwordstretchfactor}{4}
	\providecommand{\BIBentryALTinterwordspacing}{\spaceskip=\fontdimen2\font plus
		\BIBentryALTinterwordstretchfactor\fontdimen3\font minus
		\fontdimen4\font\relax}
	\providecommand{\BIBforeignlanguage}[2]{{%
			\expandafter\ifx\csname l@#1\endcsname\relax
			\typeout{** WARNING: IEEEtran.bst: No hyphenation pattern has been}%
			\typeout{** loaded for the language `#1'. Using the pattern for}%
			\typeout{** the default language instead.}%
			\else
			\language=\csname l@#1\endcsname
			\fi
			#2}}
	\providecommand{\BIBdecl}{\relax}
	\BIBdecl
	
	\bibitem{Alkhateeb2018a}
	A.~Alkhateeb, S.~Alex, P.~Varkey, Y.~Li, Q.~Qu, and D.~Tujkovic, ``Deep
	learning coordinated beamforming for highly-mobile millimeter wave systems,''
	\emph{IEEE Access}, vol.~6, pp. 37\,328--37\,348, 2018.
	
	\bibitem{Ginsburg}
	B.~P. Ginsburg, K.~Subburaj, S.~Samala, K.~Ramasubramanian, J.~Singh,
	S.~Bhatara, S.~Murali, D.~Breen, M.~Moallem, K.~Dandu \emph{et~al.}, ``A
	multimode 76-to-81{GHz} automotive radar transceiver with autonomous
	monitoring,'' in \emph{2018 IEEE International Solid-State Circuits
		Conference-(ISSCC)}.\hskip 1em plus 0.5em minus 0.4em\relax IEEE, 2018, pp.
	158--160.
	
	\bibitem{Kumari_2018}
	P.~Kumari, J.~Choi, N.~González-Prelcic, and R.~W. Heath, ``Ieee
	802.11ad-based radar: An approach to joint vehicular communication-radar
	system,'' \emph{IEEE Transactions on Vehicular Technology}, vol.~67, no.~4,
	pp. 3012--3027, 2018.
	
	\bibitem{Taha2021}
	A.~Taha, Q.~Qu, S.~Alex, P.~Wang, W.~L. Abbott, and A.~Alkhateeb, ``Millimeter
	wave {MIMO}-based depth maps for wireless virtual and augmented reality,''
	\emph{IEEE Access}, vol.~9, pp. 48\,341--48\,363, 2021.
	
	\bibitem{Alrabeiah2020}
	M.~Alrabeiah and A.~Alkhateeb, ``Deep learning for mmwave beam and blockage
	prediction using sub-6 {GHz} channels,'' \emph{IEEE Transactions on
		Communications}, vol.~68, no.~9, pp. 5504--5518, 2020.
	
	\bibitem{Ali2018}
	A.~Ali, N.~Gonz{\'a}lez-Prelcic, and R.~W. Heath, ``Millimeter wave
	beam-selection using out-of-band spatial information,'' \emph{IEEE
		Transactions on Wireless Communications}, vol.~17, no.~2, pp. 1038--1052,
	2018.
	
	\bibitem{charan2021visionposition}
	G.~Charan, T.~Osman, A.~Hredzak, N.~Thawdar, and A.~Alkhateeb,
	``Vision-position multi-modal beam prediction using real millimeter wave
	datasets,'' \emph{arXiv preprint arXiv:2111.07574}, 2021.
	
	\bibitem{Va_position}
	V.~Va, T.~Shimizu, G.~Bansal, and R.~W. Heath, ``Position-aided millimeter wave
	{V2I} beam alignment: A learning-to-rank approach,'' in \emph{2017 IEEE 28th
		Annual International Symposium on Personal, Indoor, and Mobile Radio
		Communications (PIMRC)}, 2017, pp. 1--5.
	
	\bibitem{Alrabeiah2020a}
	M.~Alrabeiah, A.~Hredzak, and A.~Alkhateeb, ``Millimeter wave base stations
	with cameras: Vision-aided beam and blockage prediction,'' in \emph{2020 IEEE
		91st Vehicular Technology Conference (VTC2020-Spring)}.\hskip 1em plus 0.5em
	minus 0.4em\relax IEEE, 2020, pp. 1--5.
	
	\bibitem{ali2019millimeter}
	A.~Ali, N.~Gonz{\'a}lez-Prelcic, and A.~Ghosh, ``Millimeter wave {V2I}
	beam-training using base-station mounted radar,'' in \emph{2019 IEEE Radar
		Conference (RadarConf)}.\hskip 1em plus 0.5em minus 0.4em\relax IEEE, 2019,
	pp. 1--5.
	
	\bibitem{Yuan_radar}
	W.~Yuan, F.~Liu, C.~Masouros, J.~Yuan, D.~W.~K. Ng, and N.~González-Prelcic,
	``Bayesian predictive beamforming for vehicular networks: A low-overhead
	joint radar-communication approach,'' \emph{IEEE Transactions on Wireless
		Communications}, vol.~20, no.~3, pp. 1442--1456, 2021.
	
	\bibitem{DeepSense}
	\BIBentryALTinterwordspacing
	A.~Alkhateeb, G.~Charan, M.~Alrabeiah, T.~Osman, A.~Hredzak, N.~Srinivas, and
	M.~Seth, ``{DeepSense 6G}: A large-scale real-world multi-modal sensing and
	communication dataset,'' \emph{available on arXiv}, 2021. [Online].
	Available: \url{https://www.DeepSense6G.net}
	\BIBentrySTDinterwordspacing
	
	\bibitem{Zhang2021}
	Y.~Zhang, M.~Alrabeiah, and A.~Alkhateeb, ``Reinforcement learning of beam
	codebooks in millimeter wave and terahertz {MIMO} systems,'' \emph{IEEE
		Transactions on Communications}, pp. 1--1, 2021.
	
	\bibitem{Li_hybrid_2019}
	X.~Li and A.~Alkhateeb, ``Deep learning for direct hybrid precoding in
	millimeter wave massive {MIMO} systems,'' in \emph{2019 53rd Asilomar
		Conference on Signals, Systems, and Computers}, 2019, pp. 800--805.
	
	\bibitem{lin2016design}
	J.-J. Lin, Y.-P. Li, W.-C. Hsu, and T.-S. Lee, ``Design of an {FMCW} radar
	baseband signal processing system for automotive application,''
	\emph{SpringerPlus}, vol.~5, no.~1, pp. 1--16, 2016.
	
	\bibitem{zhang2020object}
	G.~Zhang, H.~Li, and F.~Wenger, ``Object detection and {3D} estimation via an
	{FMCW} radar using a fully convolutional network,'' in \emph{ICASSP 2020-2020
		IEEE International Conference on Acoustics, Speech and Signal Processing
		(ICASSP)}.\hskip 1em plus 0.5em minus 0.4em\relax IEEE, 2020, pp. 4487--4491.
	
	\bibitem{dekker2017gesture}
	B.~Dekker, S.~Jacobs, A.~Kossen, M.~Kruithof, A.~Huizing, and M.~Geurts,
	``Gesture recognition with a low power {FMCW} radar and a deep convolutional
	neural network,'' in \emph{2017 European Radar Conference (EURAD)}.\hskip 1em
	plus 0.5em minus 0.4em\relax IEEE, 2017, pp. 163--166.
	
	\bibitem{kingma2014adam}
	D.~P. Kingma and J.~Ba, ``Adam: A method for stochastic optimization,''
	\emph{arXiv preprint arXiv:1412.6980}, 2014.
	
\end{thebibliography}
\end{document}